\documentclass[prl,aps,showpacs,lengthcheck,superscriptaddress,twocolumn]{revtex4}
\usepackage{amsmath}
\pdfoutput=1
\usepackage{xcolor}
\usepackage{amsfonts}
\usepackage{graphicx}
\usepackage{bm}

\newcommand{\e}{\epsilon}

\newcommand{\bw}{\mathbf{w}}
\newcommand{\xhat}{\mathbf{\hat{x}}}
\newcommand{\beq}{\begin{equation}}
\newcommand{\eeq}{\end{equation}}
\newcommand{\beqa}{\begin{eqnarray}}
\newcommand{\eeqa}{\end{eqnarray}}
\newcommand{\tr}{\tilde\rho}
\newcommand{\tg}{\tilde{\gamma}}
\newcommand{\ta}{\tilde{\alpha}}
\newcommand{\tl}{\tilde{\lambda}}
\newcommand{\tdr}{\delta\tilde{\rho}}

\begin{document}

\title{Spiral and never-settling patterns in active suspensions}

\author{X.~Yang}
\affiliation{Physics Department, Syracuse University, Syracuse NY 13244, USA}
\author{D.~Marenduzzo}
\affiliation{SUPA, School of Physics and Astronomy, University of Edinburgh, Mayfield Road, Edinburgh EH93JZ, UK}
\author{M.~C. Marchetti}
\affiliation{Physics Department and Syracuse Biomaterials Institute, Syracuse University, Syracuse NY 13244, USA}

\begin{abstract}
{We present a combined numerical and analytical study of
pattern formation in an active system where particles align, 
possess a density-dependent motility, and are subject to a logistic reaction. 
This is a model for suspensions of reproducing bacteria, but it can also 
represent, in the ordered phase,  actomyosin gels {\it in vitro} 
or {\it in vivo}. In the disordered phase, we find that motility suppression
and growth compete to yield stable or blinking patterns, which, when dense
enough, acquire internal orientational ordering, to yield asters or spirals. 
In the ordered phase, the reaction term leads to previously unobserved 
never-settling patterns which can provide a simple framework to understand 
the formation of motile and spiral patterns in actin. }

\end{abstract}

\maketitle

Bacterial suspensions self-organize into a variety of intriguing patterns  visible under the microscope. For instance,  {\it  Escherichia coli} and {\it Salmonella typhimurium} 
colonies growing on soft agar form crystalline or amorphous arrangements of high-density bacterial clumps~\cite{Murray2003,Budrene1995,Woodward1995}, 
as well as stripe patterns~\cite{Liu2011}. Biofilms exhibit even more elaborate patterns  of high density immobile regions and  low-density spots or voids~\cite{Thar2005}.
A large amount of theoretical work has been devoted to understanding the role of external chemical cues in driving such pattern formation by modeling the system via coupled nonlinear 
diffusion-reaction equations~\cite{Ben-Jacob2000,Murray2003}. It was recently demonstrated~\cite{Brenner2010,Tailleur2008,Cates2010} that effective models of reproducing organisms  
that do not explicitly include chemotaxis can also yield stable patterns, as a result of the interplay between bacterial reproduction and death and the suppression of cell motility, 
which may arise from  local crowding or biochemical signaling such as quorum sensing~\cite{Liu2011,Fu2012}.
Motility suppression has been further explored theoretically in models of self-propelled particles with density conservation~\cite{Fily2012,Redner2013,Farrell2012,Cates2012,Fu2012}. 
With only steric repulsion and no alignment, motility suppression yields  macroscopic phase separation, with large pretransitional density fluctuations~\cite{Fily2012,Redner2013}. 
In Vicsek-type models with aligning interactions~\cite{Farrell2012}, the interplay of self-trapping and alignment yields a rich collection of  traveling patterns, including bands, clumps and lanes~\cite{Farrell2012}.

In this paper we consider a continuum model of active matter where activity couples to a reactive logistic term, hence density is not conserved. This provides a model for  bacterial suspensions that incorporates cell reproduction and death~\cite{Toner2012}, motility suppression, as well as cell alignment as it may be induced  by medium-mediated hydrodynamic interaction, 
quorum sensing, or anisotropic cell shape. In the context of actin gels and solutions, a reactive term may arise due to e.g. polymerization (limited by crowding)~\cite{Whitelam2009}.

Our model is formulated in terms of two continuum fields, the  cell density, $\rho({\bf r},t)$, and  the cell polarization density, ${\bf w}({\bf r},t)$. The vector field ${\bf w}$ plays the dual role of orientational order parameter describing the local polar alignment of cells traveling in the same direction and of cell current density. The continuum equations are 
\begin{subequations}
\label{eq:hydro}
\begin{gather}
\label{rho}
\partial_t\rho=-\bm\nabla\cdot\left(v\bw-D\bm\nabla\rho\right)+\alpha\rho\left(1-\rho/\rho_s\right)\;,\\
\label{eq:w}
\begin{align}
\partial_t \bw  =&  - \left(\e- \gamma \rho\right)\bw
     - \frac{\gamma^2}{2 \epsilon}|\bw|^2 \bw  -\frac{\gamma}{4\epsilon}{\bf F}[\bw,\bm\nabla\bw]\notag\\
     & - \frac{1}{2}\bm\nabla\left(v\rho-\frac{3\gamma}{4 \epsilon}v w^2 \right) 
    +D_w\nabla^2\bw\;,
\end{align}
\end{gather}
\end{subequations}
where $v=v_0e^{-\lambda\rho}$ is the density-dependent self propulsion speed, with $\lambda$ a parameter controlling the decay of motility with increasing density, and  
${\bf F}=   \frac{1}{2}\bw \cdot \bm\nabla(v \bw) + \frac{1}{8}v\bm\nabla w^2 + \frac{3}{2}\bw \bm\nabla \cdot (v \bw) + v\bw (\nabla \cdot \bw)+v(\bw \cdot\bm\nabla)\bw$   
represents advective nonlinear terms.  The density equation is  a convection-diffusion equation, augmented by the logistic term describing bacterial growth/death at rate $\alpha$, 
with a saturation density $\rho_s$.
The polarization equation has the same structure as the well-known Toner-Tu model of flocking~\cite{Toner1995,Toner2005}, but with additional nonlinearities arising from the 
density-dependent propulsion speed (for another generic equation of polar order in active systems see also~\cite{Forest2013}). Here, $\epsilon$ is a rotational diffusion rate determined by the strength of rotational noise and $\gamma$ is the strength of the alignment interaction.
The diffusion coefficients $D$ and $D_w$ control gradients in the bacterial density and polarization. For simplicity we assume $D_w=D$.
In the following we measure  times in units of $\epsilon^{-1}$ and  lengths in units of $v_0/\epsilon$. We also define dimensionless fields 
$\tilde{\rho}=\rho/\rho_s$ and $\tilde{\bw}=\bw/\rho_s$. The model described by Eqs.~\eqref{eq:hydro} is then characterized by four dimensionless parameters, three of which are crucial 
to determine the physics: $\tilde{\lambda}=\lambda\rho_s$, $\tilde{\gamma}=\gamma \rho_s/\epsilon$, and $\tilde{\alpha}=\alpha/\epsilon$. These measure respectively the importance of 
density-dependent motility, of alignment and of death/growth. The fourth  parameter, $\tilde{D}=D\epsilon/v_0^2$  denotes the scaled magnitude of the diffusion coefficient and elastic 
constant in the polarization equation. In bacteria the birth/death rate $\alpha$ is of the order of inverse hours and the run length $v_0\epsilon$ is typically $\sim10-30\mu m$, 
with $\epsilon\sim 1-2{\rm s}^{-1}$. In this case $\ta$ is always very small and a more meaningful control parameter is $\alpha^*=\ta/\tilde{D}=\frac{\alpha v_0^2}{D\e^2}$,
which represents the ratio between pattern size and run length and can be close to $1$ in bacterial suspensions. 

\begin{figure}
\includegraphics[width=0.99\columnwidth]{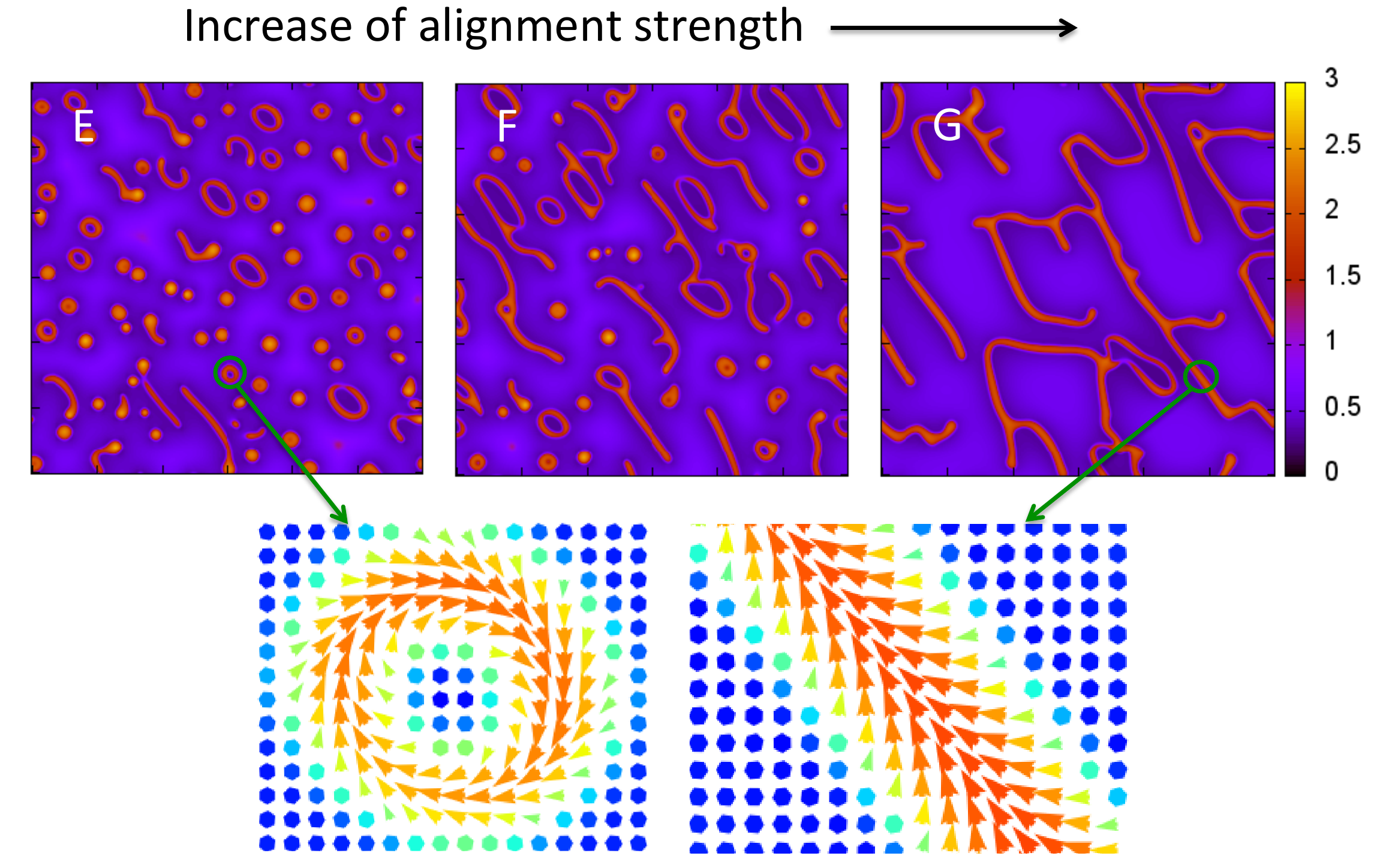}
\caption{(color online) Patterns in the polar state obtained from a uniform initial state with small random fluctuations around $\tilde{\rho}=1$ and 
$\tilde{\bw}=\bm 0$ for $\tilde{\alpha}=0.08$, $\tilde{D}=0.01$ and $\tilde{\lambda}=1.40$. Rings/lanes emerge in (E) (F) (G) at $\tilde{\gamma}=1.11, 1.25, 1.67$, respectively.}
\label{fig:polar-maps}
\end{figure}

When both motility suppression and bacteria birth/death are neglected ($\tl=\ta=0$), the equations describe the familiar Vicsek model, with a mean field transition from a homogeneous 
isotropic ($\bw=0$) state to a homogeneous polar state ($\bw =w_0\xhat$)   when  alignment  $\gamma\rho_s$ exceeds  noise $\e$, or $\tg>1$.
The polar state is also a uniformly moving state. In this model emergent structures are known to arise only in the polar state from the growth of fluctuations in the magnitude of 
polarization due to the $\gamma\rho_s$ term.
\begin{figure}
\includegraphics[width=0.89\columnwidth]{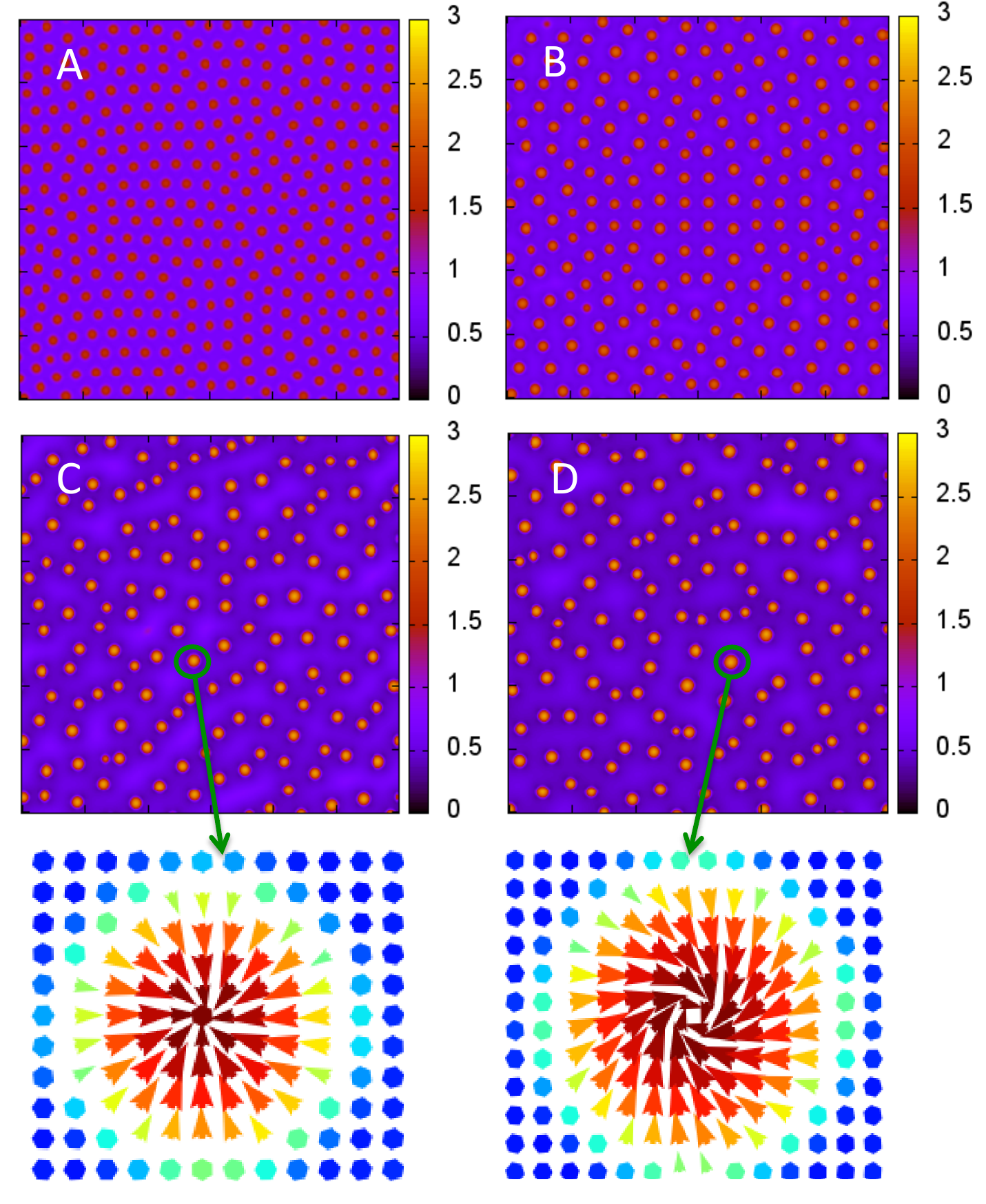}
\caption{(color online)  Color maps of the density (colorbar to the right of each snapshot) obtained by integrating the nonlinear equations numerically in a box with periodic boundary conditions and an initial isotropic 
state ($\bw=0$) of uniform density $\tr=\rho/\rho_s=1$, with small random fluctuations. All images are for $\tl=1.4$, $\tilde{D}=0.01$, $\ta=0.083$ and  (from $A$ to $D$)   
$\tg=0.50,0.58,0.75,0.95$. The high density static bacterial dots in $A$ and $B$ have zero or very small local polar order, as in~\cite{Cates2010}. In $C$ and $D$  polar order builds up 
in each dot, as highlighted by the blow ups of individual dots shown in the bottom row. Here the polarization is displayed as an arrow of length proportional to the magnitude of the 
polarization and the color indicates the density (red means high, blue low).}
\label{fig:iso-maps1}
\end{figure}
The model with no cell death and reproduction ($\ta=0$), but  with motility suppression ($\tl\not=0$), was studied  by Farrell \emph{et al} ~\cite{Farrell2012}, who showed that 
motility suppression destabilizes both the homogeneous disordered and ordered states. These authors did not, however, include cell reproduction which is essential  to explain 
length selection in experimental patterns forming on the timescales of hours or days.
In models of self-propelled particles with density conservation ($\ta=0$) and no alignment ($\tg=0$) motility suppression can yield true phase separation when
$\tl>1$~\cite{Fily2012,Redner2013}. A finite growth/death rate can arrest phase separation yielding concentric rings and high density bacterial ``dots'' not unlike those observed 
in {\em Salmonella typhimurium}~\cite{Cates2010}. We show that alignment has a number of new qualitative effects. First, the high density static dots found in~\cite{Cates2010} acquire 
characteristic polar structures of increasing complexity, such as asters and spirals, as shown in Fig.~\ref{fig:iso-maps1}. Such static polarized droplets may be relevant to bacterial 
suspensions in semisolid agar. At higher values of alignment, we find blinking dot patterns, ever-evolving rings and lanes (Fig.~\ref{fig:polar-maps})  that may be relevant 
for bacterial suspensions in liquid media, and also resemble the dynamical spiral structures  seen in the acto-myosin cytoskeleton of immobilized \emph{Dictyostelium} cells, where $\ta$ 
may describe the actin polymerization/depolymerization rate ~\cite{Whitelam2009}. 
\paragraph{Emergent patterns in the isotropic state.}
To examine the linear stability of the isotropic state for $\tg<1$, we study the dynamics of fluctuations $\delta\tilde{\rho}=\tilde{\rho}-1$ and $\Theta=\bm\nabla\cdot\tilde\bw$.
For clarity of discussion, it is useful to restore dimensionful quantities. 
Working in Fourier space, we let $(\tdr,\Theta)=\sum_{\bf q}(\rho_{\bf q} ,\Theta_{\bf q})e^{i{\bf q}\cdot{\bf r}}$.
The time evolution of $\rho_{\bf q}$ and $\Theta_{\bf q}$ is then governed by 
\begin{subequations}
\label{eq:linear}
\begin{gather}
\label{eq:rho-lin}
\partial_t\rho_{\bf q}=-v\Theta_{\bf q}-Dq^2\rho_{\bf q}-\alpha\rho_{\bf q}\;,\\
\label{eq:w-lin}
\partial_t\Theta_{\bf q}=-\e_r\Theta_{\bf q}+\frac12(v+\rho_sv')q^2\rho_{\bf q}-Dq^2\Theta_{\bf q}\;,
\end{gather}
\end{subequations}
where $\e_r=\e-\gamma\rho_s$ is the rotational diffusion rate renormalized by alignment and the prime denotes a derivative with respect to density. 
The decay/growth of fluctuations is governed  by the eigenvalues of Eqs.~\eqref{eq:linear}. The stability is controlled by the rate $s_+(q)$, given by
\begin{equation}
\label{eq:sp}
s_+(q)=-\frac{\alpha+\e_r+2Dq^2}{2}+\frac{\sqrt{(\alpha-\e_r)^2-4\e D_{sp} q^2}}{2}\;,
\end{equation}
where we have introduced an effective diffusivity 
$D_{sp}(\rho)=\left(1+\frac{1}{2}\frac{d}{d\rho}\right)\frac{v^2(\rho)}{2\e}$, with ${v}^2(\rho)/(2\e)$ the diffusivity of a single particle performing a persistent random walk with 
run speed $v(\rho)$ and diffusion (tumble) rate $\e$. In Eq.~\eqref{eq:sp} all quantities are evaluated at  $\rho_s$. For the chosen form of $v(\rho)$, the effective diffusivity 
$D_{sp}=\frac{v_0^2}{2\e}e^{-2\tl}(1-\tl)$ can change sign due to motility suppression in the region $\tl>1$. The rate $s_+$ becomes positive for 
$D_{sp}<-D\left(\sqrt{\frac{\alpha}{\e}}+\sqrt{\frac{\e_r}{\e}}\right)^2$.  
Motility suppression then promotes phase separation in regions of high and low density~\cite{Fily2012}, which is in turn arrested by the density-dependent birth/death, 
as bacteria tend to grow in low density regions and die in high density regions, hence must migrate from high density to low density regions to obtain a steady state. 
The interplay of these two mechanisms ultimately yields a stable pattern. 
For bacterial suspensions $\alpha<<\e$ and the instability condition becomes $D_{sp}<-D$. In this case the unstable mode describes the growth of density fluctuations, while polarization 
decays rapidly.  One then  recovers the limit of no alignment considered by Cates and coworkers~\cite{Cates2010}, appropriate when rotational diffusion is faster than any other process. 
In contrast, as the mean field transition is approached from below, alignment slows down the decay of polarization by decreasing its effective decay rate from $\e$ to $\e_r$, thereby 
enhancing the magnitude of the effective diffusivity and, for any fixed $\alpha$, the region of parameters where the homogeneous state is unstable (Fig.~\ref{fig:iso-PD1}).  
When $\e_r<<\alpha$, the  instability occurs for $D_{sp}<-D(\alpha/\e)$ and is  controlled by the growth  of polarization. On the other hand, for very large $\tl$ the active contribution 
to the effective  diffusivity vanishes as  $D_{sp}\rightarrow 0$, and the homogeneous state is again stable. The `knee' in the curves at $\tl=1.5$ corresponds to the minimum of $D_{sp}$. 
Finally, the instability exists even for $\alpha=0$, although in this case the system coarsens into macroscopic phase separated regions, rather than forming a stable pattern 
(see Supplementary material).  Increasing $\alpha$ for fixed alignment strength tends to stabilize the uniform state. 
\begin{figure}
\includegraphics[width=0.89\columnwidth]{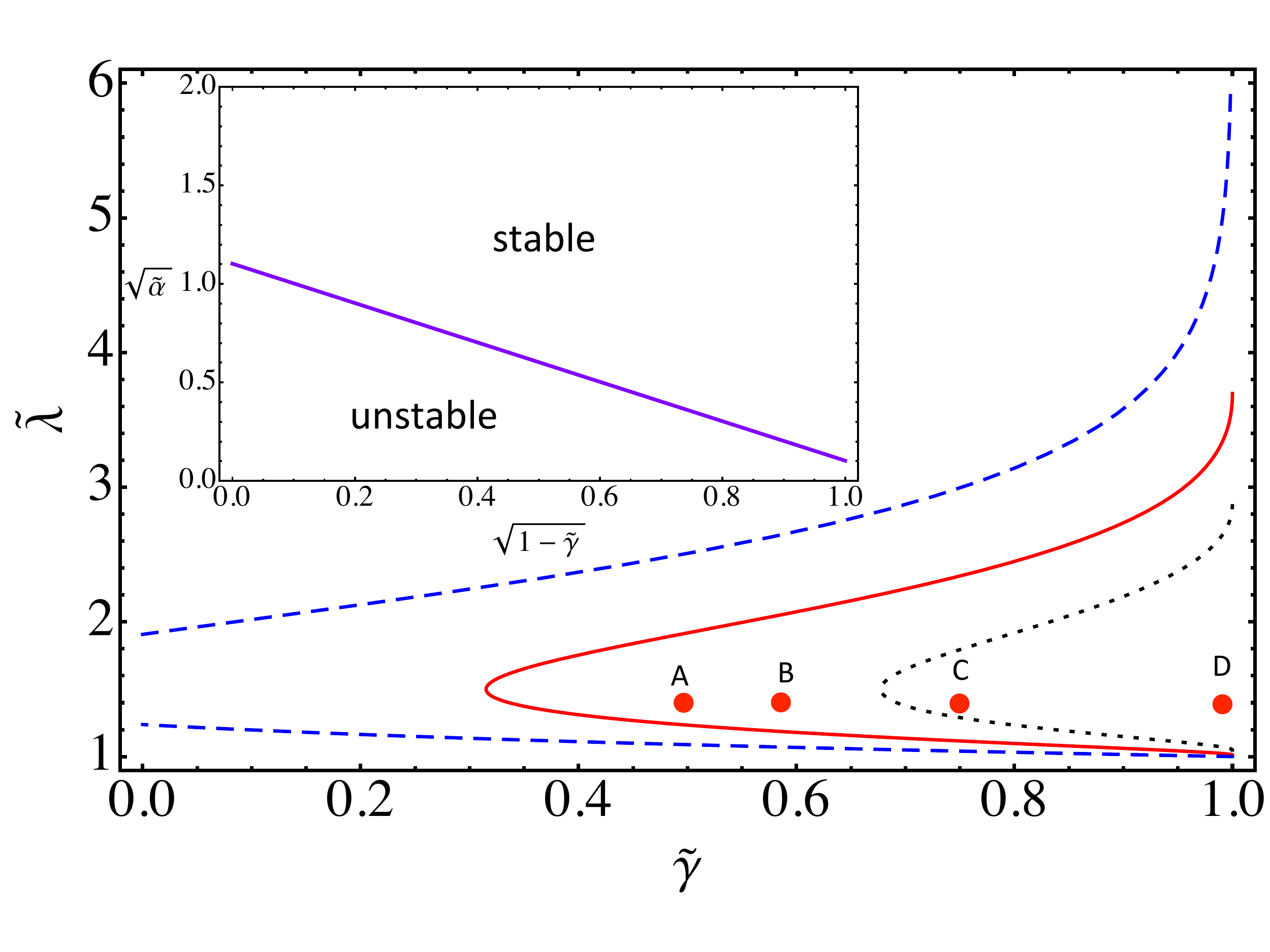}
\caption{(color online)  The  boundary of linear stability of the homogeneous state ($\tg<1$) in the $(\tl,\tg)$ plane for  
$\ta=0$ (dashed curve, blue online),  $\ta=0.083$ (solid curve, red online) and $\ta=0.30$ (dotted curve, black online). The homogeneous state is unstable in the  region to the 
right of each curve, up to the vertical axis $\tg=1$. The calculation is described in the Supplementary material.  The dots labeled A, B, C, D show the location in parameter space of 
the images shown in Fig.~\ref{fig:iso-maps1} and refer to $\alpha=0.083$. For a fixed value of $\ta$, the region of $\tl$ where the system is unstable grows as alignment increases. 
Conversely, increasing the birth/death rate for fixed $\tilde{\gamma}$ stabilizes the homogeneous state. 
This is highlighted in the inset that shows the linear stability boundary in the $(\ta,\tg)$ plane for $\tl=1.4$. Note that in the inset the horizontal axis  is $\sqrt{1-\tg}$, i.e., 
alignment increases to the left.}
\label{fig:iso-PD1}
\end{figure}

The polar structure of the dots shown in Fig.~\ref{fig:iso-maps1} arises when the density $\rho>\rho_s$ inside an individual bacterial dot is large enough that  $\gamma\rho>\epsilon$, i.e., 
the system acquires polar order. As $\tg=\gamma\rho_s/\e\rightarrow 1^-$,  the polar structure becomes even more pronounced,  with the appearance of spirals (Fig.~\ref{fig:iso-maps1}D). 
In addition, close to the transition and for a large value of the reproduction rate, the dot pattern  becomes `blinking' (see Supplementary Movie 5).  This type of blinking patterns obtained starting from 
a uniform initial  state might be observable in bacteria  in liquid media, but only if the bacterial density within the clusters achieve large enough values to induce alignment 
(this would occur, according to the Onsager theory of rods, for a volume fraction of order of the inverse aspect ratio of the bacteria, which may be $10\%$ considering flagella).
The linear stability analysis also yields the wavevector  $q_c \sim D^{-1/2}\left[\alpha\e_r\right]^{1/4}$ of the most unstable mode: the length scale $q_c^{-1}$ decreases with 
increasing $\alpha$ and $\e_r$, consistent with the behavior shown in the density maps of Fig.~\ref{fig:iso-maps1}. 

\begin{figure}
\includegraphics[width=0.99\columnwidth]{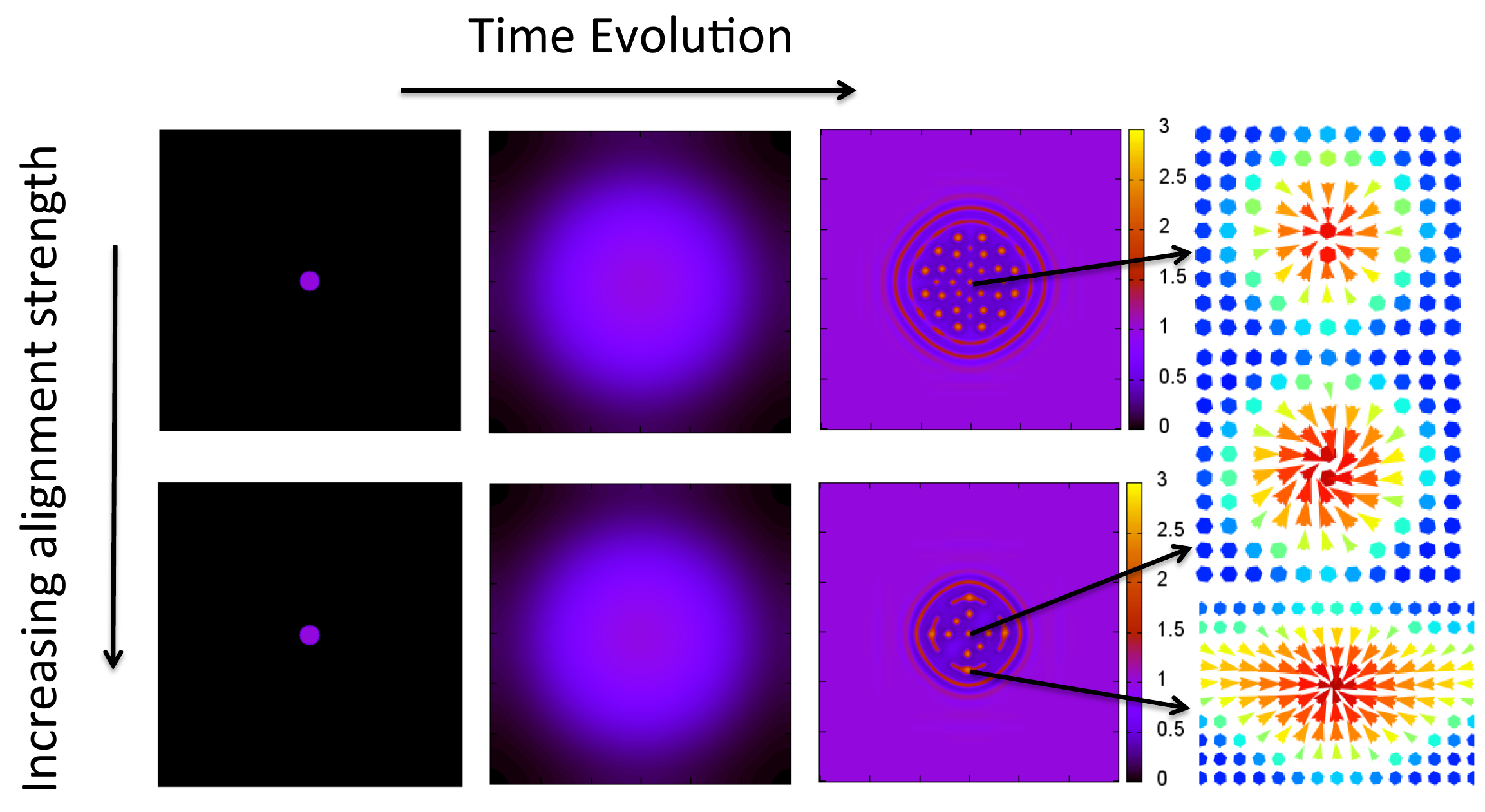}
\caption{(color online) Snapshots of the spreading of a droplet inoculated at the center of the simulation sample for $\tilde{\gamma}=0.83$ (top row) and  
$\tilde{\gamma}=0.98$ (bottom row), with $\tilde{\alpha}=0.167$, $\tilde{\lambda}=1.40$ and $\tilde{D}=0.01$. The simulation time is $1,4000,22000$ and $1,4000,14000$ 
from left to right in the unit of $\e^{-1}$ respectively. The blow-ups to the right  show the internal structure of the dots, as described in the caption of 
Fig.~\ref{fig:iso-maps1}.}
\label{fig:inoculum-iso}
\end{figure}

Finally, Fig.~\ref{fig:inoculum-iso} shows the  spreading of an initial dense bacterial droplet inoculated at the center of the simulation box. 
For large alignment, the rings and clusters become polarized and show a spiral structure. 
\paragraph{Emergent patterns in the polarized state.}
In the uniform polar state ($\bw=w_0{\bf \hat{x}}$), instabilities exist even for  $\lambda=0$ and $\alpha=0$ and these have been studied before~\cite{Bertin2009,Mishra2010}. 
The presence of the reactive birth/death term  yields new patterns. The study of the dynamics in the polarized state may not apply to usual bacterial suspensions,
where aligning effects are typically insufficient to yield an orientationally ordered phase (one exception might be provided by bacterial swarms). Such a study is instead highly relevant 
for actomyosin systems, where (i) immobilized motors ({\it in vitro}, e.g. in standard motility assays~\cite{Schaller2010})  or treadmilling ({\it in vivo}) lead to propulsion; (ii) steric effects may cause motility suppression; and 
(iii) polymerization (limited by actin crowding) may be, in the simplest approximation, added on as a reaction term~\cite{Whitelam2009}~\footnote{Our choice corresponds to a linear term 
for polymerization and a quadratic saturation term mimicking steric effects -- the work in Ref.~\cite{Whitelam2009} employs a different nonlinear reaction, where polymerization is quadratic. }.

For the parameters used in our model,  when $\lambda=\alpha=0$ fluctuations in the magnitude of polarization of wave vector ${\bf q}\Vert x$ destabilize the uniform state 
for $1\leq\tg\leq 11/7$, while splay fluctuations with ${\bf q}\perp x$ are always unstable. Both these longitudinal and transverse instabilities have been discussed extensively. 
In particular, the longitudinal instability has been argued to signal the onset of high density ordered bands normal to the direction of mean polarization traveling in a 
disordered low density background~\cite{Bertin2009,Mishra2010}. The  suppression of motility induced by a finite  $\lambda$ yields a host of complex structures, including traveling dots, 
stripes and lanes that coarsen at long times  into anisotropic phase separated states~\cite{Farrell2012}. A finite value of $\ta$ is again needed to arrest the phase separation and 
yield stable patterns with a characteristic length scale, as shown in Fig.~\ref{fig:polar-maps}. In the region $\tg>1$ the patterns are always dynamical, with the rings continuously 
breaking up and reconnecting, as displayed in  movie 3 of the Supplementary material. Such never-settling states are, interestingly, virtually absent (except for simple 
traveling waves) in previous versions of the model, neglecting either alignment or growth. 

The dynamical patterns in Fig.~\ref{fig:polar-maps}  may qualitatively explain the existence of and transition between dynamical spiral patterns  observed in the cytoskeleton of 
Dictyostelium~\cite{Whitelam2009}. 
Here individual cells were first immobilized by depolimerizing actin with latrunculin. Upon latrunculin reduction, static spots of actin were seen to form in the cortex. At later times, 
the spots become dynamical and eventually turn into spiral  waves that closely resemble our ever-evolving rings, and allow the cell to eventually resume motility. 
The experimental observation was modeled in~\cite{Whitelam2009} by a reaction-diffusion model  where actin density and orientation growth are controlled by a chemical inhibitor. 
Our model provides a simpler interpretation of the self-organization where chemical signaling does not need to be explicitly incorporated in the dynamics, but only enters through effective 
parameters such as $\lambda$ and $\alpha$, with the latter representing here the actin polymerization/depolymerization rate. 

In summary, we have studied  an active system where constituents (i) align, (ii) move at a density-dependent speed and (iii) reproduce, or undergo a logistic reaction. The first model 
system this case refers to is that of a bacterial suspension.
While it had been demonstrated before that the interplay of motility suppression and bacterial growth/death can yield stable structures in a simple effective model even in the absence of 
chemotaxis~\cite{Cates2010}, our work shows, for the first time, that alignment interactions play a major role in pattern selection.
When alignment is small ($\tg<<1$) polarization decays on the time scale of the inverse rotational diffusion rate $\e$, that in bacteria is much faster than reproduction. 
Bacterial organization is then controlled entirely by the dynamics of the density and results in isotropic patterns of high density dots, depending on the initial conditions, 
as seen before. Larger values of the alignment interaction ($\tg\leq 1$) slow down the decay of polarization fluctuations. In addition, although the mean density $\rho_s$ may be below the 
value $\e/\gamma$ required for the onset of mean-field polar order, the local density in the bacterial spots may exceed this value and the dots acquire polar structure of increasing 
complexity (asters and spirals) even for $\gamma\rho_s<\e$. Above the mean-field transition to a polar state ($\tg>1$) the patterns become dynamical, with ever-evolving rings and travelling 
lanes that have not been seen before. These structures resembles the actin spirals that have been observed in the cytoskeleton of Dictyostelium~\cite{Whitelam2009}.  
It might be interesting to study the effect of choosing different forms for the reaction term, other than a simple logistic.

Our work thus demonstrates that effective models are capable of yielding a rich dynamics and to generate the large variety of patterns seen in experiments~\cite{Brenner2010}.
An important open challenge remains the derivation of such effective models from microscopic ones, providing an understanding of how the effective parameters are determined by the underlying mechanical and chemical mechanisms.

MCM and XY were supported by the National Science Foundation through awards DMR-1004789 and DGE-1068780.
\bibliography{References}
\end{document}